# Flexomagnetoelectric effect in Sr$_2$IrO$_4$ thin films

Xin Liu[1,2,3#], Ting Hu[4#], Yujun Zhang[5], Xueli Xu[6], Biao Wu[1,2], Zongwei Ma[6], Peng Lv[7], Yuelin Zhang[1,2], Shih-Wen Huang[8], Jialu Wu[9], Jing Ma[9], Jiawang Hong[7], Zhigao Sheng[6], Chenglong Jia[10*], Erjun Kan[4*], Ce-Wen Nan[9*] and Jinxing Zhang[1,2*]

[1]Department of Physics, Beijing Normal University, Beijing 100875, China

[2]Key Laboratory of Multiscale Spin Physics, Ministry of Education, China

[3]SwissFEL, Paul Scherrer Institute, Villigen PSI 5232, Switzerland

[4]School of Science, Nanjing University of Science and Technology, Nanjing 210094, China

[5]Institute of High Energy Physics, Chinese Academy of Sciences, Beijing 100049, China

[6]Anhui Province Key Laboratory of Condensed Matter Physics at Extreme Conditions, High Magnetic Field Laboratory of the Chinese Academy of Science, Hefei 230031, China

[7]School of Aerospace Engineering, Beijing Institute of Technology, Beijing 100081, China

[8]Swiss Light Source, Paul Scherrer Institute, Villigen PSI 5232, Switzerland

[9]School of Materials Science and Engineering, Tsinghua University, Beijing 100084, China

[10]Key Laboratory for Magnetism and Magnetic Materials of the Ministry of Education and Lanzhou Center for Theoretical Physics, Lanzhou University, 73000, Lanzhou, China

E-mail: cljia@lzu.edu.cn; ekan@njust.edu.cn; cwnan@tsinghua.edu.cn; jxzhang@bnu.edu.cn

Symmetry engineering is explicitly effective to manipulate and even create phases and orderings in strongly correlated materials. Flexural stress is universally practical to break the space-inversion or time-reversal symmetry. Here, by introducing strain gradient in a centrosymmetric antiferromagnet $Sr_2IrO_4$, the space-inversion symmetry is broken accompanying a non-equivalent O $p$-Ir $d$ orbital hybridization along $z$ axis. Thus, emergent polar phase and out-of-plane magnetic moment have been simultaneously observed in these asymmetric $Sr_2IrO_4$ thin films, which both are absent in its ground state. Furthermore, upon the application of magnetic field, such polarization can be controlled by modifying the occupied $d$ orbitals through spin-orbit interaction, giving rise to a flexomagnetoelectric effect. This work provides a general strategy to artificially design multiple symmetries and ferroic orderings in strongly correlated systems.

In condensed matters, the space-inversion and time-reversal symmetries almost determine all of the electrical and magnetic properties of solids. For example, break of time-reversal symmetry creates spin ordering, while break of space-inversion symmetry gives rise to polar ordering [1,2]. The coupling between polar and spin orderings, namely magnetoelectric (ME) effect, could bring diverse promising applications ranging from information processing, sensing to energy harvesting etc [1-3]. However, among the 122 kinds of Shubnikov-Heesch point groups, only 13 groups can exhibit spontaneous polarization and magnetization simultaneously in a single phase (point groups: 1, 2, 2', m, m', 3, 3m', 4, 4m'm', m'm2', m'm'2, 6, 6m'm') [4,5]. Due to such symmetry restrictions of polar (space-inversion symmetry) and spin (time-reversal symmetry) orderings [1,2,4-7], there are quite limited materials showing the magnetoelectric effect in nature [5]. Therefore, it is desirable to develop advanced strategies in order to achieve symmetry design for creating ferroic orderings and ME effect at nano- or atomic-scale, especially applicable to a variety of thin-film materials which could be beneficial to the future device integration [3,5,8].

Flexural stress can give rise to an inhomogenous strain that may break space-inversion symmetry in all 32 crystalline point groups [9-13]. In recent decades, a variety of flexural-induced phases in ferroics (e.g. ferroelectrics or ferromagnets) [14-18] and -mediated phenomena/functionalities (e.g. pyroelectricity, photovoltaic and caloric effects, etc.) have been discovered [13,19-21]. In general, the inhomogenous strain can be achieved in epitaxial nanoscale thin films due to the strain relaxation when the thickness exceeds the critical value [22]. In the strain-relaxed epitaxial thin films, a large strain gradient ($\sim 10^6$ m$^{-1}$) can be obtained, which is six or even seven orders of magnitude larger than the one in the flexural bulks [13,22]. Therefore, the strain gradient in epitaxial thin films may be effective to control materials' symmetries and therefore ME effect at nanometric scales (as seen in Fig. 1).

In this letter, a centrosymmetric $Sr_2IrO_4$ (SIO) with a strong spin-orbit coupling (SOC) (~0.4 eV) [23-26] was selected as a model system to demonstrate that the flexural stress or strain gradient is a general strategy for simultaneously designing space-inversion and time-reversal symmetries. Owing to the intrinsic spin-orbit interaction and engineered symmetry break, a strong non-equivalent $pd$ hybridization occurs, which is considered as a key energy term to bridge the emergent polar and spin orderings [3,27], leading to a flexomagnetoelectric effect.

High-quality epitaxial SIO thin films on SrTiO$_3$ (100) (STO) substrates have been grown by laser molecular beam epitaxy with *in-situ* reflection high-energy electron diffraction (RHEED) (Supplemental Material Methods and Supplemental Material Fig. S1). A 45° rotation along $z$-axis of SIO on STO was verified by X-ray diffraction (XRD) $\varphi$ scan (Supplemental Material Methods) as seen in Supplemental Material Fig. S2, which is consistent with results reported elsewhere [28]. X-ray reciprocal space mapping (Supplemental Material Methods) was shown in Supplemental Material Fig. S3, further confirming the epitaxial growth of the high-quality SIO thin films from 60 nm to 500 nm without any impurity phase. Due to the lattice mismatch between SIO and STO substrate (the in-plane lattice mismatch is about 0.45% and SIO suffers an in-plane tensile strain), strain relaxation forms when thickness increases as schematically depicted in Fig. 2(a) [29], which can be indicated by the broadened and shifted structural peak (1118) of SIO thin films in Supplemental Material Fig. S3. Figure 2(b) summarizes the out-of-plane average strain of SIO thin films (blue dots), which are -1.17%, -0.77%, -0.37%, -0.35% and -0.31% for the 60 nm, 120 nm, 200 nm, 300 nm and 500 nm films, respectively. In order to study the relaxation-induced strain gradient, Williamson-Hall (W-H) plots [30] from the XRD $\theta$-$2\theta$ results (Supplemental Material Methods) of SIO thin films have been analyzed in Supplemental Material Fig. S4. The estimated strain gradients as a function of SIO thicknesses are shown in Fig. 2(b) (pink dots), where the maximum one (~$1.05\times10^6$ m$^{-1}$) occurs at ~300 nm. For the 500 nm film, a fully strain relaxation results in a little decrease of the strain gradient [29].

Large strain gradient is effective to beak the space-inversion symmetry [31,32] and may modify the electronic structures of the SIO thin films. Firstly, temperature-dependent dielectric constants were collected along the z direction of SIO thin films (Supplemental Material Methods) as shown in Fig. 2(c). Dielectric anomalies were observed at ~20 K in 200 nm, 300 nm and 500 nm films, but not in 60 nm and 120 nm films, implying a possible polar disorder-order phase transition [33] in the SIO with a large strain gradient, or named as flexo-SIO. According to the Neumann principle [34], the electric tensor components will be affected by the long-range polar ordering, which can be detected by the second harmonic generation (SHG). By measuring SHG signals (Supplemental Material Methods and Supplemental Material Fig. S5) from the flexo-SIO (~500 nm thin film), an emergent out-of-plane polar ordering (SHG signal in P) was observed at 15 K as seen in Fig. 2(d). However, for the 120 nm SIO thin film without a large strain gradient, no out-of-plane SHG signal was observed (see Supplemental Material Fig. S6). Furthermore, via synchrotron X-ray diffraction (Supplemental Material Methods), a Bragg reflection of SIO (1 2 15) was measured as a function of temperature in the flexo-SIO (Supplemental Material Fig. S7). This Bragg peak satisfies the reflection condition $h + k$ = odd and $l$ = odd, which is sensitive to the distortion of oxygen octahedra [24,35]. As seen in Fig. 2(e), an abrupt enhancement of the peak intensity was observed when the polar phase transition occurred (~20 K). Such a dramatic change of oxygen octahedra indicates a first-order structural transition corresponding to the emergence of the polar phase in the flexo-SIO.

This emergent polar phase has been understood by symmetry analysis. Given that the point group of the bulk SIO is 4/$m$ [25], when introducing the strain gradient, this centrosymmetric structure can be broken and the point group changes from 4/m to 4 due to the lacking of mirror symmetry [36] along the z direction, producing an asymmetric length of Ir-O bond ($\delta_L$ = Ir-O$_u$ − Ir-O$_d$ ≠ 0, $\delta_L$ is the flexural-stress- or strain-gradient-induced change of out-of-plane Ir-O$_{u/d}$ bond length, where O$_u$ and O$_d$ denote the apical and bottom oxygens in the O-Ir-O chain) as seen in Figs. 3(a) and 3(b). This point group 4 is non-centrosymmetric and belongs to the polar point groups [4,37], providing the prerequisite for the polar ordering along the z-direction.

Density functional theory (DFT) calculations (Supplemental Material Methods) were carried out to further explore the microscopic origin of this emergent polar ordering in a flexo-SIO with SOC. According to calculations of the density of state (DOS) in Supplemental Material Fig. S8, enhanced DOSs and hybridizations of O $p_y$ orbitals and Ir $d_{yz}$ orbitals are observed when introducing the strain gradient in SIO. In particular, the difference of DOS between $O_u$ and $O_d$ ($O_u - O_d$) $p_y$ orbitals are plotted in Figs. 3(c) and 3(d). In the flexo-SIO, a pronounced enhancement of DOS was observed in $O_u - O_d$ $p_y$ orbitals, giving rise to asymmetric electronic DOS and non-equivalent hybridizations along the z-direction. Considering the involved $t_{2g}$ in the state $\Gamma_7$ near the Fermi level and such an enhanced $pd$ hybridization (overlapped in the green shadow in Figs. 3(c) and 3(d)) in the flexo-SIO, the Hamiltonian can be expressed as (Supplemental Material Methods) [38]:

$$H_{pd} = V_{pd\pi}[d_{zx}^+ p_{u,x} + d_{yz}^+ p_{u,y} - d_{zx}^+ p_{d,x} - d_{yz}^+ p_{d,y}] + H.c. \quad (1)$$

where $d$ and $u$ represent down and up orbitals respectively, $H.c.$ is Hermitian adjoint. According to the symmetry analysis and DFT calculations, the asymmetric Ir-$O_{u/d}$ bond leads to a non-equivalent hybridization, which will induce non-zero residual value ($\delta_I = I_u - I_d \neq 0$) from the unbalanced overlap integral $I_u$ and $I_d$, where $I_i = \langle d_{yz}|z|p_{i,y}\rangle$. Thus, given that the shapes of involved $pd$ orbitals, as schematically shown in Fig. 3(e), we have the polarization developing along the z direction [39]:

$$P_z \approx \min\left(\frac{\lambda}{V}, 1\right)\frac{V}{\Delta}\delta_I m_z^2 \quad (2)$$

where $\lambda$ is SOC, $V$ is energy of hybridization between $p$ and $d$ orbitals, $\Delta$ is charge transfer energy between $p$ and $d$ orbitals, $m_z$ is the local magnetic moment of IrO$_6$ octahedra along z-direction, the integral $\delta_I$ is approximately estimated as $\delta_I \approx \frac{16}{27} Z_O^{5/2} Z_{Ir}^{7/2} / \left(\frac{Z_O}{2} + \frac{Z_{Ir}}{2}\right)^6 \delta_L e a_0$ [40], where $\delta_L$ is the flexural-stress- or strain-gradient-induced change of out-of-plane Ir-$O_{u/d}$ bond length, $a_0$ is Bohr radius, $e$ is the electron charge, $Z_O/Z_{Ir}$ is the Clementi-Raimondi effective charges of O/Ir. As expressed in Eq. (2), the non-zero residual value ($\delta_I$), SOC ($\lambda$) and z-direction magnetic moment ($m_z$) are key elements for the emergent polarization.

Apart from $\delta_I$ and $\lambda$, our DFT calculation further indicates that $z$-direction magnetic moments ($m_z$) may exist in the SIO with a large strain gradient (see Supplemental Material Table I), which should be responsible for the emergent polar phase according to the derived Eq. (2). In order to experimentally verify this, we characterized the temperature-dependent magnetization $M$ ($T$) (Supplemental Material Methods) in the SIO thin films. Both 120-nm and 300-nm-thick SIO thin films show in-plane magnetic moments as seen in Fig. 4(a). In surprise, an unconventional $z$-direction magnetic moment was observed in the 300-nm-thick thin film as shown in Fig. 4(b), which provides the direct evidence of $m_z$ in the flexo-SIO. While for the 120-nm-thick SIO thin film, no $z$-direction magnetic moment was observed (Fig. 4(b)), which is similar to the case of single-crystal SIO without an inhomogenous strain [26,41]. The verified $z$-direction magnetic moments (schematically in Fig. 4(c)) further support the non-equivalent $pd$ hybridization theory as described in Eq. (2).

This emergent polarization exhibits the potential to couple with the magnetic moment $m_z$. Therefore, the ME effect was explored by dynamic measurements (Supplemental Material Methods and Supplemental Material Fig. S9). A maximum ME output along the $z$-direction was obtained near the polar phase transition temperature (~20 K) of the flexo-SIO thin films with thicknesses of 200 nm, 300 nm and 500 nm, where the ME coefficients are ~52 mV/cm·Oe, ~56 mV/cm·Oe and ~25 mV/cm·Oe (in Fig. 5(a)) respectively. There is no ME signal observed in 60 nm and 120 nm thin films (no $m_z$ observed). In the flexo-SIO, the effective magnetic field on Ir along $z$-direction couples to both the $z$-component of spins ($m_z$) and the $d$ orbital via spin-orbit interaction. External magnetic field could perturbate the spins along $z$-direction and manipulate the $z$-component effective magnetic field. The occupied states of $d$ orbitals can be deformed by this effective magnetic field through strong spin-orbit interaction to produce the ME effect [27,42] as seen in Fig. 5(b). Within the linear ME response region,

the ME coefficient can be expressed as:

$$\alpha = dP_z/dH = 2\min\left(\frac{\lambda}{V}, 1\right)\frac{V}{\Delta}\delta_I m_z dm_z/dH \qquad (3)$$

where the $dm_z/dH$ is the magnetic susceptibility, $\delta_I$ is the flexural-stress- or strain-gradient-induced non-zero residual value. Approximately, such a strain-gradient-induced ME coefficient can be estimated in the range of 10~100 mV/cm·Oe (Supplemental Material Methods), which has the same order of magnitude with our experimental results. Thus, we define the observed ME coupling in the flexo-SIO as a flexomagnetoelectric effect.

In summary, via introducing a strain gradient in SIO thin films, an emergent polar phase and unconventional magnetic moment along $z$-direction were discovered. We propose a non-equivalent $pd$ hybridization theory to explain the microscopic origin of the observed polar and spin orderings. This polar ordering couples with spin ordering of Ir by strong spin-orbit interaction to produce a flexomagnetoelectric effect. Such a strategy of symmetry design is universal to centrosymmetric materials with a strongly electron correlation for creating ferroic orderings and may even triggering more emergent quantum phenomena.

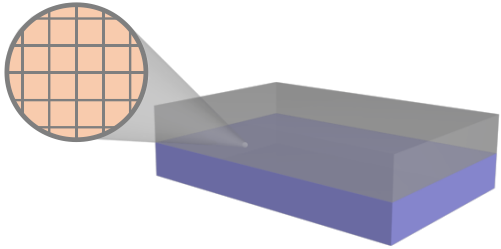 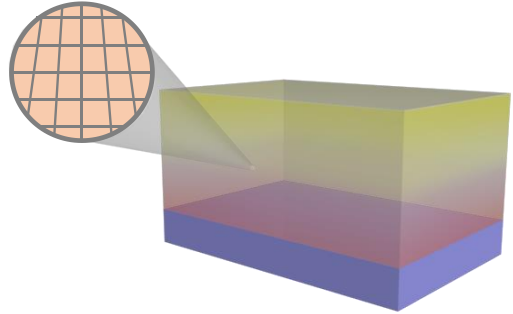

FIG. 1. Symmetry engineering by a large strain gradient. (a) Schematic of epitaxial thin film on the substrate with homogeneous strain, persisting the inversion symmetry. (b) Schematic of epitaxial thin film on the substrate with a strain gradient, breaking the inversion symmetry, which might induce the emergent orderings and phases.

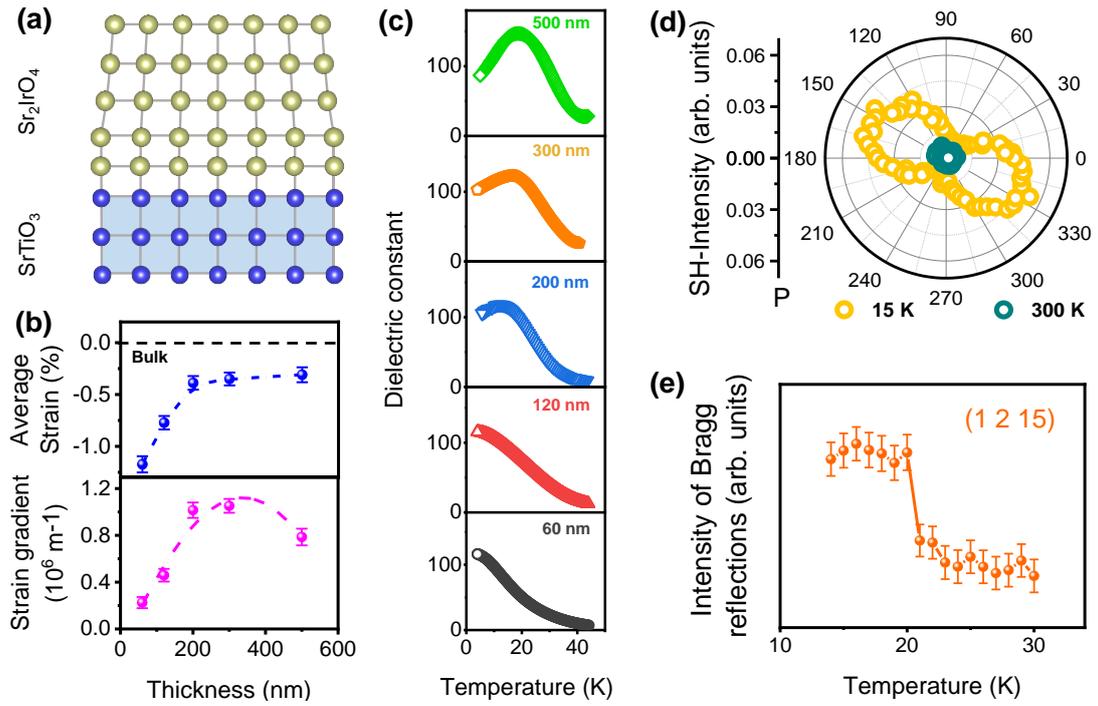

FIG. 2. Emergent polar phase in flexo-Sr$_2$IrO$_4$ (SIO). (a) Schematic of strain gradient of SIO on SrTiO$_3$ (STO) substrate. (b) Top: Average strain in SIO thin film. Bottom: Estimated out-of-plane strain gradient (pink dots) as function of SIO thickness. The maximum strain gradient is estimated at ~300 nm. (c) Temperature dependence of dielectric constant along $z$ direction for various thickness of SIO thin films, dielectric anomaly was observed for 200-, 300- and 500-nm-thick samples. (d) Second harmonic generation of 500-nm-thick SIO thin film along P, indicating a polar ordering along $z$ direction at 15 K. (e) Temperature-dependent intensity of (1 2 15) Bragg reflections corresponding to the distortion of the oxygen octahedra near the polar phase transition.

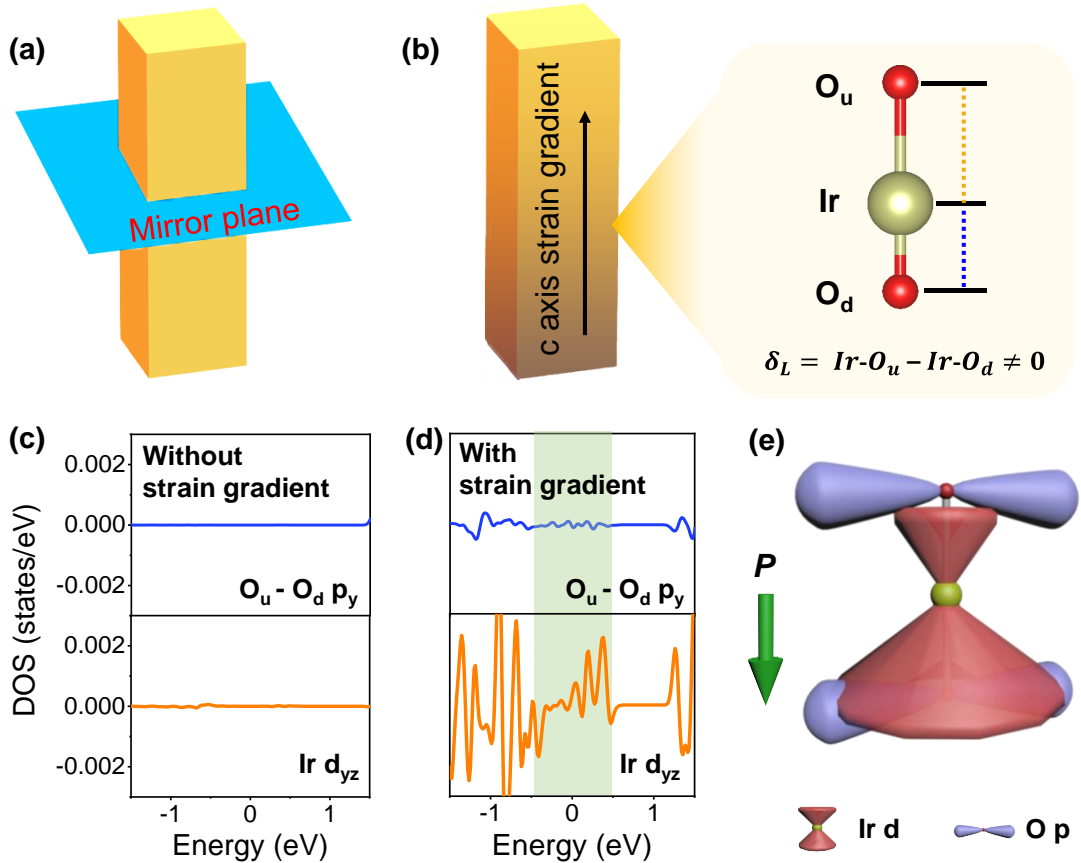

FIG. 3. Origin of the emergent polar phase. (a),(b) Symmetry analysis of SIO without and with strain gradient. The point group of bulk SIO belongs to $4/m$, where one of the mirror planes exist in this centrosymmetric structure. For SIO thin films under strain gradient, the centrosymmetric structure will be broken to the point group of 4 (one of polar point groups) due to lacking of this mirror plane with an asymmetric length of Ir-O bond ($\delta_L = $ Ir-$O_u$ − Ir-$O_d \neq 0$, $\delta_L$ is the strain-gradient-induced change of out-of-plane Ir-$O_{u/d}$ bond length, where $O_u$ and $O_d$ demonstrate the apical and bottom oxygen in the O-Ir-O chain). (c),(d) Density of states of $O_u$ - $O_d$ $p_y$ and Ir $d_{yz}$ orbitals without (left) and with (right) strain gradient. The non-equivalent hybridization of $O_u$ - $O_d$ $p_y$ and Ir $d_{yz}$ orbitals is enhanced near the Fermi level in flexo-SIO (Light green shadow). (e) Schematic of the $pd$ hybridization induced polarization in flexo-SIO with SOC.

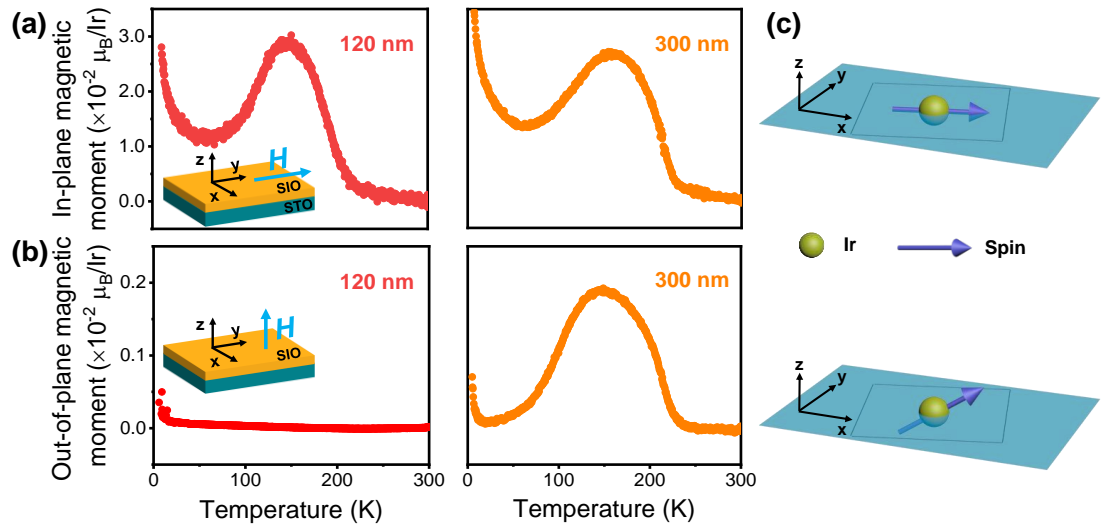

FIG. 4. Unconventional out-of-plane magnetism in flexo-SIO. Magnetic field of 1000 Oe was applied after the zero-field-cooling process. (a) In-plane magnetic moment as a function of temperature in 120-nm and 300-nm-thick SIO thin films. Inset shows the schematic of the measurements. (b) Out-of-plane magnetic moment as a function of temperature in 120-nm and 300-nm-thick SIO thin films. Inset shows the schematic of the measurements. (c) Top: spin configuration in fully strained SIO, bottom: spin configuration in flexo-SIO.

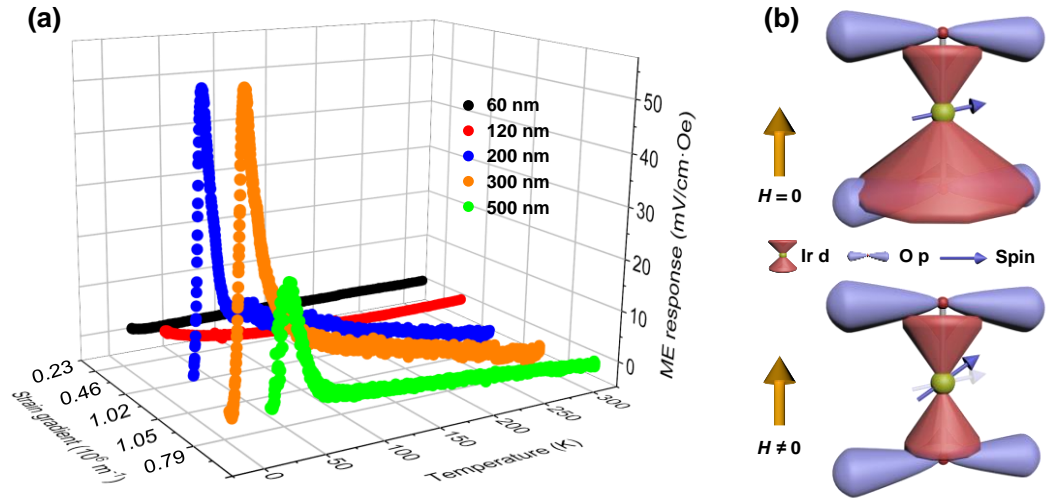

FIG. 5. Flexomagnetoelectric effect. (a) Magnetoelectric effect in SIO thin films, where the maximum magnetoelectric outputs occur near the temperature of polar phase transition of the flexo-SIO. (b) Schematic of the ME effect in flexo-SIO. The external magnetic field perturbate the spins along $z$ direction to modify the occupied $d$ orbitals by spin-orbit interaction, thus modify the polarization, producing ME effect.

## Author contributions

J.Z. and X.L. designed the experiments and prepared the manuscript. X.L. prepared the thin films, performed the dielectric and dynamic ME response measurements. X.L. and Y.Z did the analysis of strain gradient. X.X., Z.M, B.W. and Z.S. conduct the SHG experiment and analysis. X.L. and S.-W.H. did the synchrotron X-ray diffraction at Swiss Light source. X.L., J.W., Y.J.Z., J.M. and C.-W.N. did the RSM and SQUID measurements. X.L., P.L., Y.Z., and J.H. did the symmetry analysis. T.H. and E.K. carried out the DFT calculations. C.J. provided the theoretical model. All authors were involved in the discussion of the experimental and theoretical results and the revision of the manuscript.

## Acknowledgements

This work was supported by the National Natural Science Foundation of China (No.52225205, No.12174164, No.91963201), the Ministry of Science and Technology of China (No. 2023YFA1406500), the National Key R&D Program of China through Contract No. 2021YFA0718700, the Fundamental Research Funds for the Central Universities and the 111 Project under Grant No. B2006. We gratefully acknowledge the support from the Materials Science beamline (X04SA) of the Swiss Light Source in Paul Scherrer Institute and the beamline 1W1A of the Beijing Synchrotron Radiation Facility.

## Competing interests

The authors declare no competing interests.

# Supplemental Material

# Flexomagnetoelectric effect in $Sr_2IrO_4$ thin films

Xin Liu[1,2,3#], Ting Hu[4#], Yujun Zhang[5], Xueli Xu[6], Biao Wu[1,2], Zongwei Ma[6], Peng Lv[7], Yuelin Zhang[1,2], Shih-Wen Huang[8], Jialu Wu[9], Jing Ma[9], Jiawang Hong[7], Zhigao Sheng[6], Chenglong Jia[10*], Erjun Kan[4*], Ce-Wen Nan[9*] and Jinxing Zhang[1,2*]

[1]Department of Physics, Beijing Normal University, Beijing 100875, China

[2]Key Laboratory of Multiscale Spin Physics, Ministry of Education, China

[3]SwissFEL, Paul Scherrer Institute, Villigen PSI 5232, Switzerland

[4]School of Science, Nanjing University of Science and Technology, Nanjing 210094, China

[5]Institute of High Energy Physics, Chinese Academy of Sciences, Beijing 100049, China

[6]Anhui Province Key Laboratory of Condensed Matter Physics at Extreme Conditions, High Magnetic Field Laboratory of the Chinese Academy of Science, Hefei 230031, China

[7]School of Aerospace Engineering, Beijing Institute of Technology, Beijing 100081, China

[8]Swiss Light Source, Paul Scherrer Institute, Villigen PSI 5232, Switzerland

[9]School of Materials Science and Engineering, Tsinghua University, Beijing 100084, China

[10]Key Laboratory for Magnetism and Magnetic Materials of the Ministry of Education and Lanzhou Center for Theoretical Physics, Lanzhou University, 73000, Lanzhou, China

E-mail: cljia@lzu.edu.cn; ekan@njust.edu.cn; cwnan@tsinghua.edu.cn; jxzhang@bnu.edu.cn

**Section 1: Methods**

**Sample preparations.**

Via the laser-MBE with *in-situ* reflection high-energy electron diffraction (RHEED), high quality epitaxial SIO thin films were synthesized on STO (001) substrates with the thickness of 60, 120, 200, 300 and 500 nm, respectively. All the films were grown under an oxygen pressure of 0.02 mbar at 800 °C with the heating rate of 20 °C/min. A KrF excimer laser with a wavelength of 248 nm was used with the laser fluence of ~1 J cm$^{-2}$. The repetition rate was 2 Hz. After deposition, the films were slowly cooled (5 °C/min) under an oxygen pressure of 100 mbar. For the samples in transport measurements, a bottom electrode of LaNiO$_3$ was deposited, in order to maintain the same strain gradient, the thickness of bottom electrode is ~6 nm fully strained by the substrate.

**Reciprocal space mapping (RSM), *θ-2θ* and *φ* scans.**

The RSM and *θ-2θ* scans were characterized using an x-ray diffractometer (X'pert Pro2, PANalytical) with Cu Kα1 radiation (λ= 1.5406 Å). The *φ* scan was carried out at the beamline 1W1A of the Beijing Synchrotron Radiation Facility.

**Dielectric and second harmonic generation (SHG) measurements.**

The temperature dependence of dielectric constant was measured along the *z* direction by Precision LCR Meter (Keysight E2980 AL) equipped with cryostats (PPMS DynaCool; Quantum Design). The AC voltage was 0.1 V at 10 kHz during the dielectric measurement.

The SHG measurements were carried out at High magnetic field laboratory of the Chinese Academy of Science Hefei. A 45° reflection geometry was used with a fundamental wavelength of 800 nm. A half-wave plate was used to rotate the polarization angle of the incident pump pulses and a Glan prism was used to rotated the polarization angle of the output SH pulses. The SH photons could be selected by the monochrometer and be transformed by a photomultiplier tube which was recorded by a lock-in amplifier.

**Synchrotron X-ray diffraction.**

The temperature dependence of Bragg reflections was measured by synchrotron X-ray diffraction at MS beamline (X04SA) of Swiss Light Source. The vertical geometry was used with the energy of 11.18 keV.

**DFT calculations.**

DFT calculations were performed using the planewave based projector augmented wave (PAW) [1,2] method as implemented in the Vienna ab initio simulation package (VASP) [3] within the generalized gradient approximation (GGA) [4] including spin-orbit coupling (SOC). The kinetic energy cutoff of the plane wave basis was chosen to be 400 eV. The convergence criteria for electronic relaxations are $10^{-5}$ eV per atom. For a proper description of the electronic correlation effects, we have included Hubbard correction [5] to the $5d$ orbitals of Ir. The values of $U$ and $J$ for Ir are 2 eV and 0.2 eV, respectively [6]. At each fixed strain gradient, the $c$-coordinates of all atoms are adjusted according to the strain gradient and fixed during the calculations.

**Theory model of $pd$ hybridization.**

In this model, the polarization is proportional to $\min(\lambda/V, 1)(V/\Delta)I$, where $\lambda$ is SOC, $V$ is energy of hybridization between $p$ and $d$ orbitals, $\Delta$ is charge transfer energy between $p$ and $d$ orbitals, $I$ is the overlap integral [7,8]. In SIO, because of the on-site SOC in Ir, the degenerated $t_{2g}$ orbitals split into a fully occupied $J_{\text{eff}}$=3/2 quartet ($\Gamma_8$) and a singly occupied upper $J_{\text{eff}}$ =1/2 doublet ($\Gamma_7$) spin-orbit coupled states, which is further splitted by Coulomb interaction $U$ to generate the band gap [9]. The twofold degenerate higher energy states $\Gamma_7$ near the Fermi level in the strong SOC limit can be expressed as [10,11]:

$$\left|J=\tfrac{1}{2}, m=-\tfrac{1}{2}\right\rangle = -\tfrac{1}{\sqrt{3}}|d_{xy\downarrow}\rangle + \tfrac{1}{\sqrt{3}}|d_{yz\uparrow}\rangle - \tfrac{i}{\sqrt{3}}|d_{zx\uparrow}\rangle \quad (1)$$

$$\left|J=\tfrac{1}{2}, m=+\tfrac{1}{2}\right\rangle = +\tfrac{1}{\sqrt{3}}|d_{xy\uparrow}\rangle + \tfrac{1}{\sqrt{3}}|d_{yz\downarrow}\rangle + \tfrac{i}{\sqrt{3}}|d_{zx\downarrow}\rangle \quad (2)$$

in which, however, the $d_{xy}$ orbital does not hybridize with any of the oxygen $p$

orbitals in the linear O-Ir-O geometry along the $z$-direction. This $d_{xy}$ orbital will be ignored in the following discussion. Then the Hamiltonian can be expressed as [12]:

$$H_{pd} = V_{pd\pi}[d_{zx}^+ p_{u,x} + d_{yz}^+ p_{u,y} - d_{zx}^+ p_{d,x} - d_{yz}^+ p_{d,y}] + H.c. \qquad (3)$$

where $d$ and $u$ represent down and up orbitals respectively, $H.c.$ is Hermitian adjoint.

Due to the shapes of involved $d$ and $p$ orbitals, the polarization along the $z$-direction can only exist when the residual value $\delta_I = I_u - I_d$ is non-zero, where $I_u = \langle d_{yz}|z|p_{u,y}\rangle$ and $I_d = \langle d_{yz}|z|p_{d,y}\rangle$ are up and down overlap integrals respectively. According to the symmetry analysis and DFT calculations in flexo-SIO, the mirror symmetry along the $z$-direction is broken with an asymmetric Ir-O$_{u/d}$ bond ($\delta_L$ = Ir-O$_u$ − Ir-O$_d$ ≠ 0), which will induce non-zero residual value $\delta_I$. $\delta_I$ can be approximately expressed as $\delta_I \approx \frac{16}{27} Z_O^{5/2} Z_{Ir}^{7/2} / \left(\frac{Z_O}{2} + \frac{Z_{Ir}}{2}\right)^6 \delta_L e a_0$ [8], where $a_0$ is Bohr radius, $e$ is the electron charge, $\delta_L$ is the strain-gradient-induced change of out-of-plane Ir-O$_{u/d}$ bond length and $Z_O/Z_{Ir}$ is the Clementi-Raimondi effective charges of O/Ir. In addition, considering this non-zero residual value along O-Ir-O bond, the $z$-direction magnetic moment should exist according to the $pd$ hybridization theory, then the polarization can be expressed as $P_z \approx \min\left(\frac{\lambda}{V}, 1\right) \frac{V}{\Delta} \delta_I m_z^2$ [7], where $m_z$ is the local magnetic moment of IrO$_6$ octahedra along $z$-direction, $\delta_I$ is non-zero residual value induced by strain gradient.

Taking $\delta_L \approx 1\%$ from DFT results, we obtain the ferroelectric polarization along the $z$ direction $P_z \approx 10~\mu C/m^2$. For the ME coupling coefficient $\alpha = dP/dH = 2\min\left(\frac{\lambda}{V}, 1\right) \frac{V}{\Delta} \delta_I m_z dm_z/dH$, where the $dm_z/dH$ is the magnetic susceptibility. Taking the magnetic susceptibility in the range of $4\pi \times (10^{-5} \sim 10^{-4})$, then the coefficient can be estimated in the range of 10~100 mV/cm·Oe.

**_M(T)_ measurements.**

Temperature dependence of magnetic moment was obtained by a SQUID

magnetometer (Quantum Design). After zero-field cooling, a magnetic field of 1000 Oe was applied in the *xy*-plane or along *z*-axis to detect the magnetic moment.

**Dynamic ME effect measurements.**

Supplemental Material Fig. S9 depicts the schematic picture of the experimental configurations for ME response measurement. An alternating magnetic field (~2.5 Oe) is generated by AC current (Keithley 6221 AC source), and the ME-induced AC voltage on the samples was collected by a lock-in amplifier (Stanford Research SR830). The AC magnetic field is along the *z*-direction. Coaxial cables were used in all the measurements.

**Section 2: Additional Figs. S1-S9 and Table I**

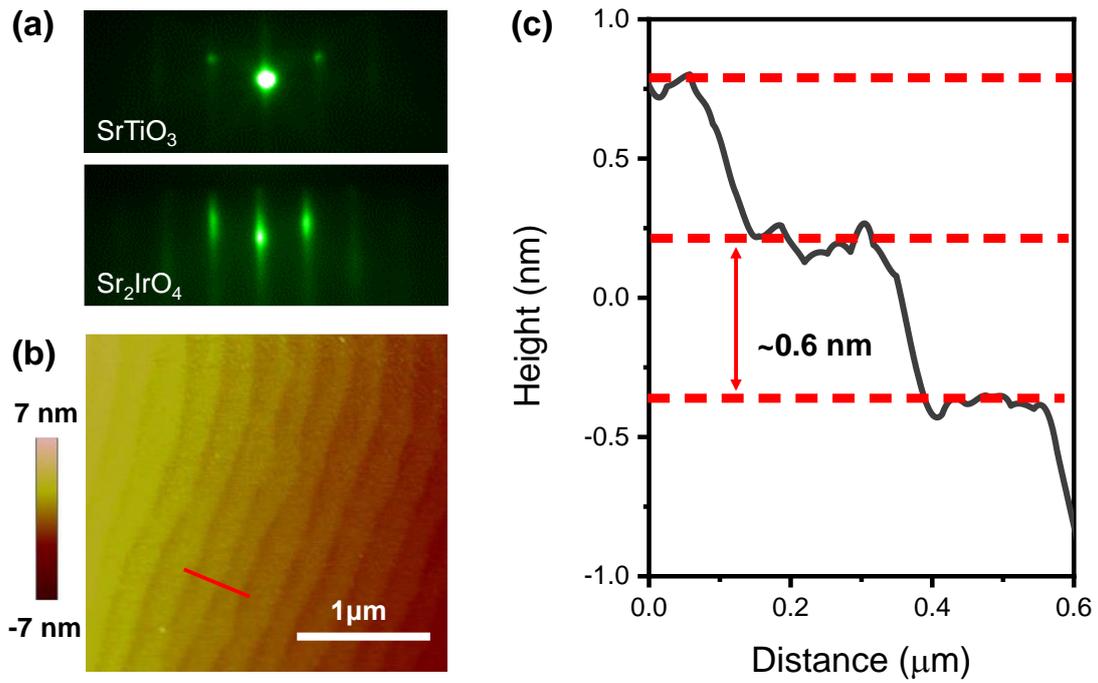

FIG. S1. (a) RHEED pattern of STO substrate and SIO thin film during the epitaxial growth, demonstrating an atomically flat surface. (b) Topography of 60-nm-thick SIO thin film with the step height of ~0.6 nm in (c).

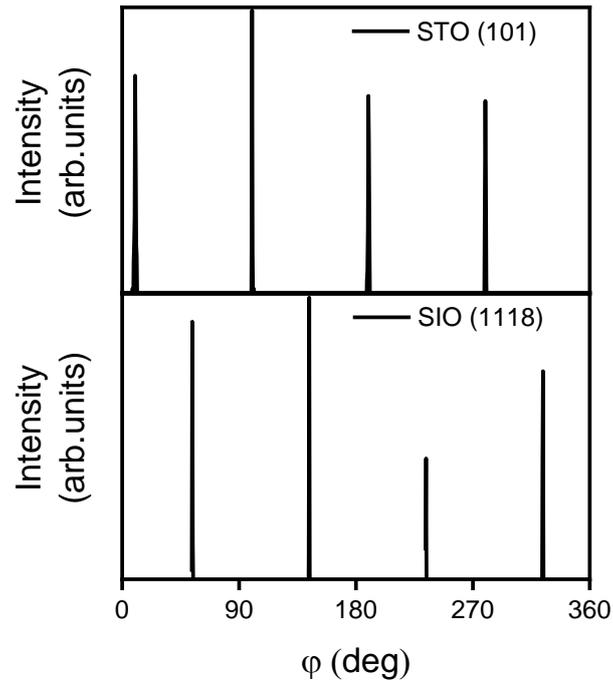

FIG. S2. $\varphi$ scan of SIO (1118) and STO (101), indicating a 45° rotation along $z$ direction of SIO epitaxially grown on STO (100) substrate.

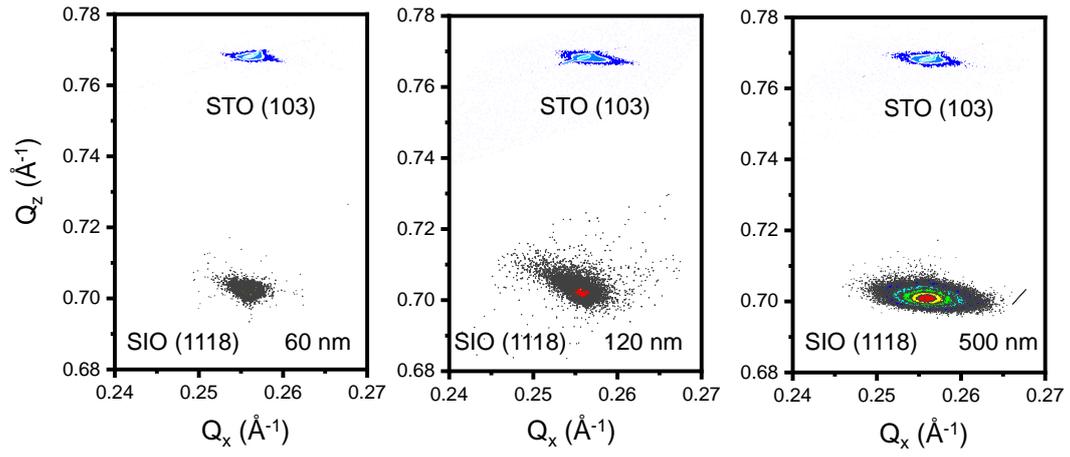

FIG. S3. X-ray reciprocal space mapping for SIO thin films with the thickness of 60 nm, 120 nm and 500 nm respectively.

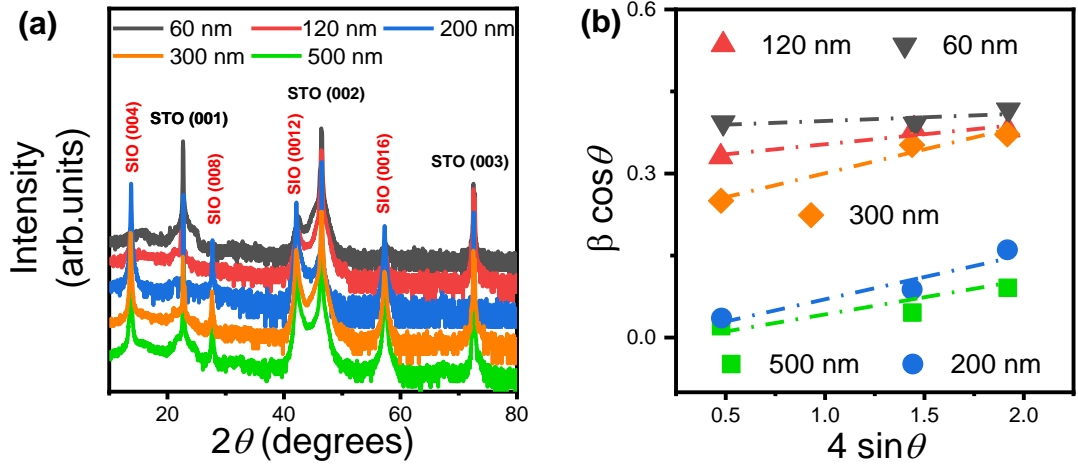

FIG. S4. (a) X-ray diffraction of $\theta/2\theta$ scan for a $c$-oriented $Sr_2IrO_4$ thin film with various thickness. (b) The out-of-plane Williamson-Hall plots for the inhomogeneous strain ($\varepsilon_I$) of the $Sr_2IrO_4$ films. Three peaks were selected from XRD data to obtain this out-of-plane change of strain. The inhomogeneous strain was determined from the slope of the fitting equation [13,14]:

$$\beta \cos\theta = \frac{K\lambda_w}{D} + 4\varepsilon_I \sin\theta$$

where $\beta = \beta_{measured} - \beta_{instrument}$, and $\beta_{measured}$ is the measured line width of the peaks in (a), $\beta_{instrument}$ is estimated from the peak width $\beta_{substrate}$ of a nearby substrate peak [13], $\lambda_w$ is the wavelength of this X-ray, $K$ is the geometrical constant and $D$ is the coherence length. According to the fitting in (b), we obtained the slop ($\varepsilon_I$): 1.35%, 3.67%, 8.12%, 8.41% and 6.27% for 60 nm, 120 nm, 200 nm, 300 nm and 500 nm respectively. Thus, the estimated strain gradients were calculated as $0.23 \times 10^6$ m$^{-1}$, $0.46 \times 10^6$ m$^{-1}$, $1.02 \times 10^6$ m$^{-1}$, $1.05 \times 10^6$ m$^{-1}$ and $0.79 \times 10^6$ m$^{-1}$ [14].

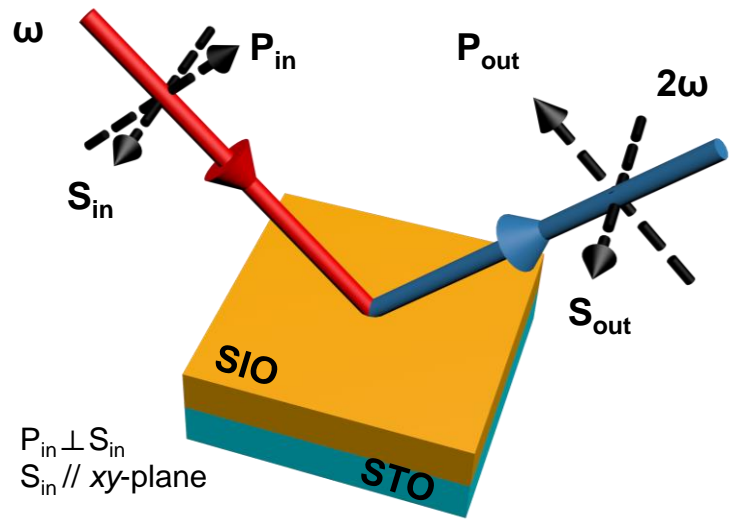

FIG. S5. Schematic picture of the experimental configurations for SHG measurement.

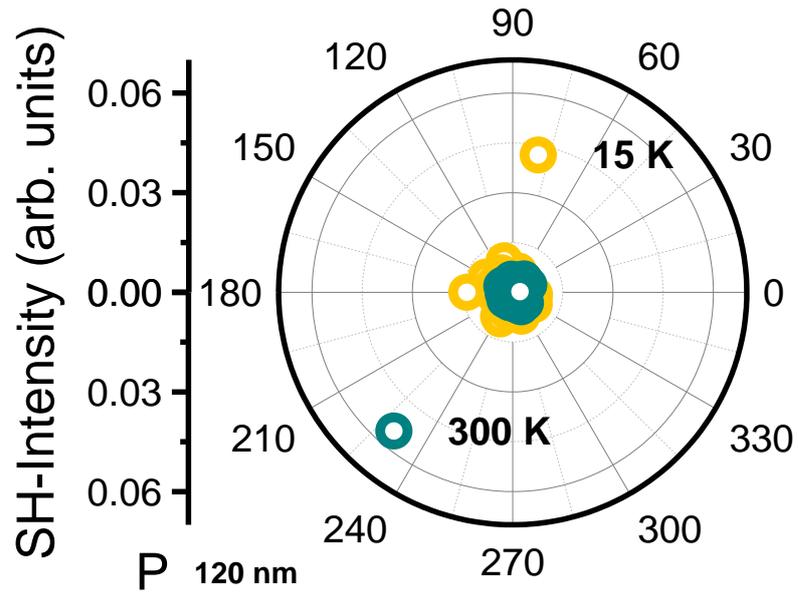

FIG. S6. Second harmonic generation of 120-nm-thick SIO thin film. No out-of-plane polarization is observed.

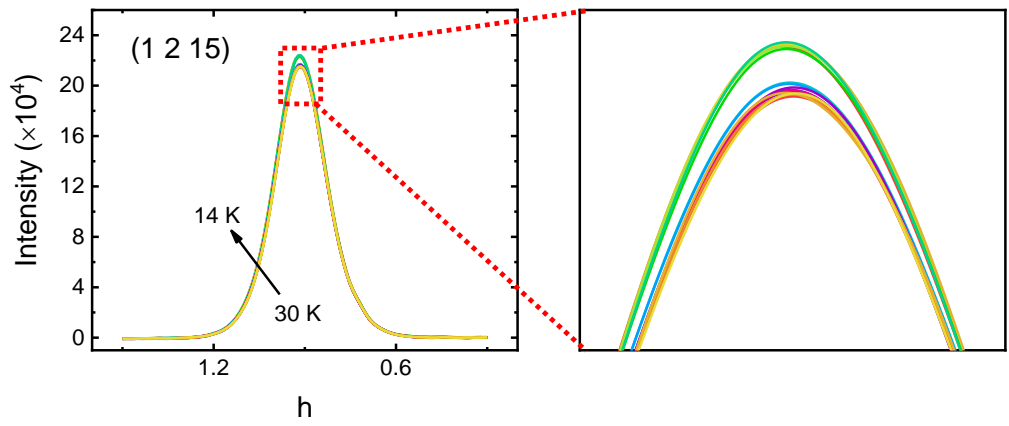

FIG. S7. Temperature dependent measurement of Bragg peak (1 2 15) in flexo-SIO. The reduced *h* value indicates the strain relaxation.

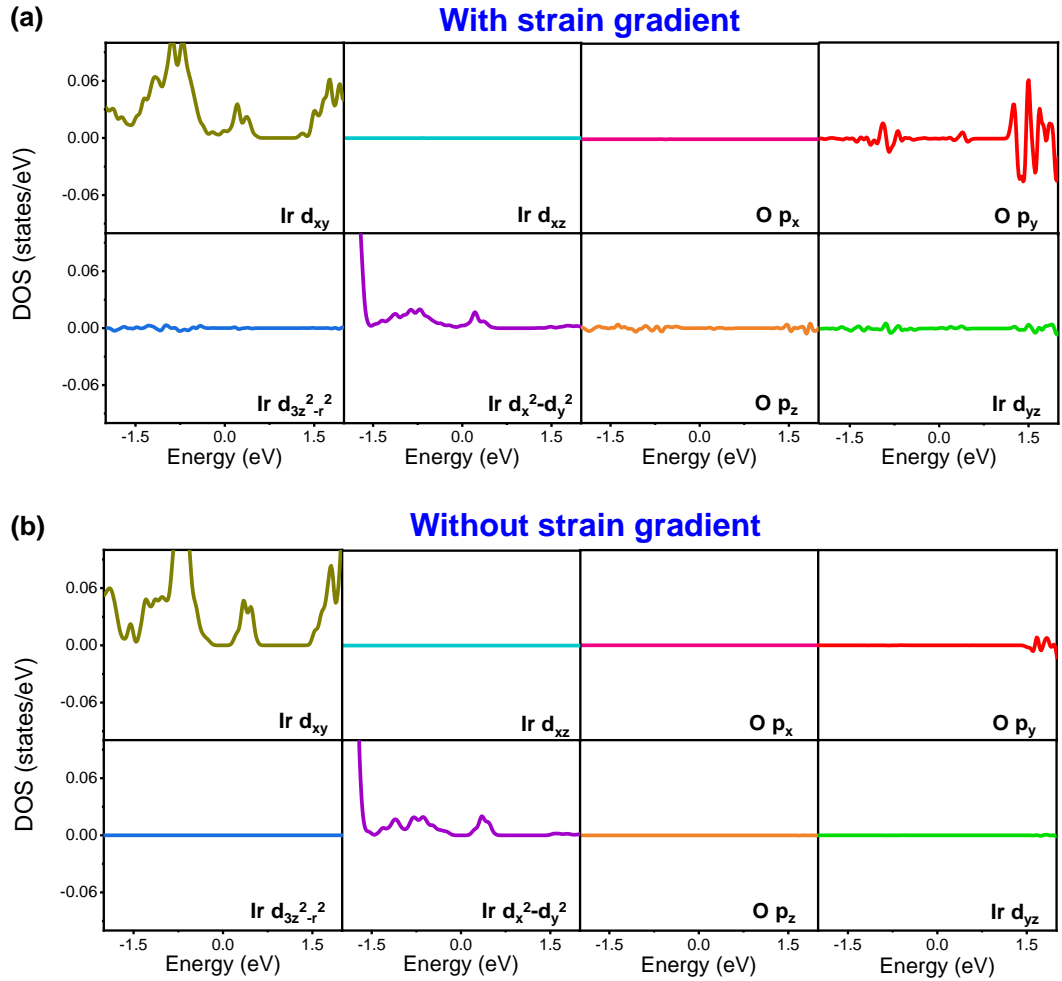

FIG. S8. Density of states of O $p$ and Ir $d$ orbitals in SIO with (a) and without (b) strain gradient. The pronounced enhanced O $p_y$ and Ir $d_{yz}$ orbitals are plotted in Figs. 3(c) and 3(d) for in-depth discussion.

Table I Magnetic moments along different axis from DFT

| Without strain gradient | | | With strain gradient | | |
|---|---|---|---|---|---|
| $m_x$ (μ$_B$) | $m_y$ (μ$_B$) | $m_z$ (μ$_B$) | $m_x$ (μ$_B$) | $m_y$ (μ$_B$) | $m_z$ (μ$_B$) |
| -0.030 | 0.132 | 0.000 | 0.026 | -0.077 | 0.002 |
| -0.033 | -0.131 | -0.000 | 0.024 | 0.077 | -0.002 |
| -0.044 | 0.146 | 0.000 | -0.032 | 0.088 | 0.004 |
| -0.044 | -0.146 | -0.000 | -0.033 | -0.088 | -0.004 |
| 0.041 | 0.145 | 0.000 | -0.039 | -0.112 | -0.000 |
| 0.041 | -0.145 | -0.000 | -0.039 | 0.112 | 0.000 |
| 0.028 | 0.134 | -0.000 | 0.031 | 0.007 | 0.047 |
| 0.025 | -0.134 | 0.000 | 0.028 | -0.005 | -0.047 |

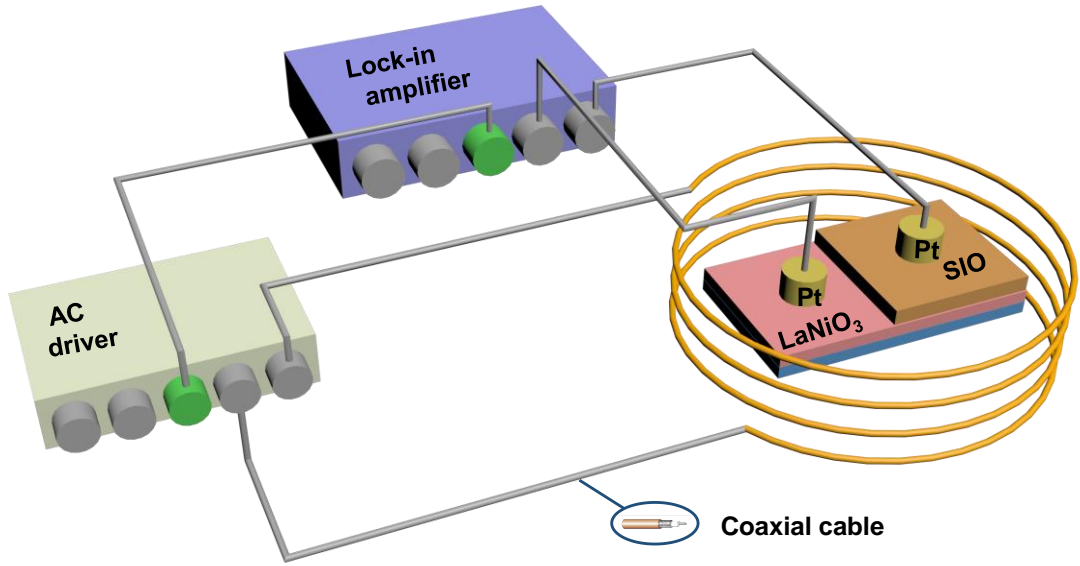

FIG. S9. The schematic picture of the experimental configurations for dynamic ME response measurement.